\documentclass{aa}

\usepackage{graphicx}
\usepackage{txfonts}
\begin{document}

   \title{Formation of complex organic molecules in molecular clouds – acetaldehyde, vinyl alcohol, ketene, and ethanol via the "energetic" processing of C$_2$H$_2$ ice
        }
\titlerunning{“energetic” processing on C$_2$H$_2$ ice}
   \author{K.-J. Chuang\inst{1,2}, G. Fedoseev\inst{2,3,4}, C. Scir\`{e}\inst{3}, G. A. Baratta\inst{3}, C. Jäger\inst{1}, Th. Henning\inst{5}, 
        H. Linnartz\inst{2}, and M. E. Palumbo\inst{3}          
    }
\authorrunning{Chuang et al.}
   \institute{
                                Laboratory Astrophysics Group of the Max Planck Institute for Astronomy at the Friedrich Schiller University Jena, Institute of Solid State Physics,       Helmholtzweg 3, D-07743 Jena, Germany \email{chuang@mpia.de}
              \and
              Laboratory for Astrophysics, Leiden Observatory, Leiden University, P.O. Box 9513, NL-2300 RA Leiden, the Netherlands
              \and
              INAF – Osservatorio Astrofisico di Catania, via Santa Sofia 78, 95123 Catania, Italy
              \and
              Research Laboratory for Astrochemistry, Ural Federal University, Kuibysheva St. 48, 620026 Ekaterinburg, Russia
              \and
              Max Planck Institute for Astronomy, Königstuhl 17, 69117 Heidelberg, Germany 
              }

  \date{}

 
  \abstract
   {The simultaneous detection of organic molecules of the form C$_2$H$_{\text{n}}$O, such as ketene (CH$_2$CO), acetaldehyde (CH$_3$CHO), and ethanol (CH$_3$CH$_2$OH), toward early star-forming regions offers hints of a shared chemical history. Several reaction routes have been proposed and experimentally verified under various interstellar conditions to explain the  formation pathways involved. Most noticeably, the non-energetic processing of C$_2$H$_2$ ice with OH-radicals and H-atoms was shown to provide formation routes to ketene, acetaldehyde, ethanol, and vinyl alcohol (CH$_2$CHOH) along the H$_2$O formation sequence on grain surfaces in translucent clouds.}
   {In this work, the non-energetic formation scheme is extended with laboratory measurements focusing on the energetic counterpart, induced by cosmic rays penetrating the H$_2$O-rich ice mantle. The focus here is on the H$^+$ radiolysis of interstellar C$_2$H$_2$:H$_2$O ice analogs at 17 K.}
   {Ultra-high vacuum experiments were performed to investigate the 200 keV H$^+$ radiolysis chemistry of predeposited C$_2$H$_2$:H$_2$O ices, both as mixed and layered geometries.  Fourier-transform infrared spectroscopy was used to monitor in situ newly formed species as a function of the accumulated energy dose (or H$^+$ fluence). The infrared (IR) spectral assignments are further confirmed in isotope labeling experiments using H$_2$$^{18}$O.}
   {The energetic processing of C$_2$H$_2$:H$_2$O ice not only results in the formation of (semi-) saturated hydrocarbons (C$_2$H$_4$ and C$_2$H$_6$) and polyynes as well as cumulenes (C$_4$H$_2$ and C$_4$H$_4$), but it also efficiently forms O-bearing COMs, including vinyl alcohol, ketene, acetaldehyde, and ethanol, for which the reaction cross-section and product composition are derived. A clear composition transition of the product, from H-poor to H-rich species, is observed as a function of the accumulated energy dose. Furthermore, the astronomical relevance of the resulting reaction network is discussed.}
   {}


\maketitle{}
\section{Introduction}   
Interstellar complex organic molecules (COMs), which are referred to as organic compounds consisting of more than six atoms,  are present in various star-forming stages, stretching from interstellar clouds to comets in our Solar System \citep{Herbst2009, Biver2014, Altwegg2017, Herbst2017}. Aside from a large number of  heterogeneous COMs being hosted in massive hot cores and giant molecular clouds, such as Orion KL, Sgr B2N, and TMC-1,   over the past decade, several oxygen-bearing COMs have also been unambiguously identified toward low-mass protostars, such as IRAS 16923-2422, NGC1333 IRAS 2A, and NGC1333 IRAS 4A, as well as prestellar sources, such as B1-b, L1689B, and B5 \citep{Bisschop2008, Oberg2010, Bacmann2012, Cernicharo2012, Coutens2015, Taquet2015, Jorgensen2016, Taquet2017, Rivilla2017}. For example, COMs described by the formula C$_2$H$_{\text{n}}$O$_2$ (n=4 and 6), such as glycolaldehyde (HCOCH$_2$OH), methyl formate (HCOOCH$_3$), and ethylene glycol (HOCH$_2$CH$_2$OH), have been observed toward the solar-mass protostar IRAS 16923-2422 \citep{Jorgensen2012, Jorgensen2016}. The simultaneous detection of these COMs, characterized by their degree of hydrogenation, implies that they are likely to share similar formation mechanisms. Such a chemical link is also found in another group of O-bearing COMs expressed by the formula C$_2$H$_{\text{n}}$O (n=2, 4, and 6), namely, ketene (CH$_2$CO), acetaldehyde (CH$_3$CHO), and ethanol (CH$_3$CH$_2$OH). Also these species have been detected toward several protostellar sources, such as NGC 7129 FIRS 2, SVS13-A, IRAS 16923-2422, and the L1157-B1 shock region \citep{Bisschop2007, Fuente2014, Lefloch2017, Bianchi2018, Jorgensen2020}.

Unlike the case of molecules of the form C$_2$H$_{\text{n}}$O$_2$, the formation chemistry of species of the form C$_2$H$_{\text{n}}$O, is still under debate \citep{Enrique-Romero2016, Enrique-Romero2020}. Several solid-state mechanisms, such as radical associations, O-atom additions, and hydroxylation, or gas-phase routes have been proposed \citep{Charnley2004, Balucani2015, Taquet2016, Chuang2020}. For example, \cite{Bennett2005a} reported acetaldehyde formation through radical association reactions between CH$_3$ and HCO, which immediately form upon the electron bombardment of a CO:CH$_4$ ice mixture. Furthermore, electron-induced chemistry using isotope-labeled reactants in combination with mass spectrometry confirmed the formation of other derivatives, including ketene and vinyl alcohol \citep{Maity2014, Abplanalp2016}. More recently, the reaction of CH$_3$ + HCO was studied theoretically, suggesting that the branching ratios of the final products, that is, CH$_3$CHO and CO + CH$_4$, are dependent on the initial orientation of the reactants \citep{Lamberts2019}. Another possible solid-state pathway leading to C$_2$H$_{\text{n}}$O species is through the interactions of hydrocarbons, such as ethylene (C$_2$H$_4$) and ethane (C$_2$H$_6$), with suprathermal ("hot") O-atoms \citep{DeMore1969}. For example, \cite{Hawkins1983} reported the formation of three different C$_2$H$_4$O isomers (i.e., acetaldehyde, vinyl alcohol, and ethylene oxide) and ketene upon photolysis of C$_2$H$_4$:O$_3$ in an Ar matrix at 15-20 K. Similar products (with exception of ketene) have also been found in electron-induced suprathermal O-atom experiments using a C$_2$H$_4$:CO$_2$ ice mixture \citep{Bennett2005b}. Recent laboratory research by \cite{Bergner2019} studying the interactions between C$_2$H$_4$ and excited O-atoms ($^1$D) generated upon UV-photolysis of CO$_2$ only confirmed the formation of acetaldehyde and ethylene oxide, while vinyl alcohol formation was not reported. Their experimental results also showed that the excited O-atoms ($^1$D) reacting with C$_2$H$_2$ or C$_2$H$_6$ results in ketene and in acetaldehyde and ethanol, respectively. These previous studies indicate that the production of particular C$_2$H$_{\text{n}}$O species formed in suprathermal O-atom induced chemistry is strongly dependent on the type of initial reactants and the excited state of the reacting O-atom. Moreover, H-atom abstraction reactions induced by O-atoms ($^1$D or $^3$P) and the recombination of H- and O-atoms originating from energetic dissociation are expected to produce hydroxyl (OH) radicals, which in turn can further react with unsaturated hydrocarbons (e.g., C$_2$H$_2$ and C$_2$H$_4$) through relatively low barriers \citep{Miller1989, Zellner1984, Basiuk2004, Cleary2006, McKee2007}. In laboratory studies, \cite{Hudson1997} reported the C$_2$H$_2$ hydroxylation resulting in the formation of acetaldehyde and ethanol in H$^+$ radiolysis of a C$_2$H$_2$:H$_2$O ice mixture at 15 K. Later, other chemically relevant species such as vinyl alcohol and ketene were also identified in similar ice analogs \citep{Hudson2003, Hudson2013}. Similar organic products have also been observed upon UV-photon irradiation of a C$_2$H$_2$:H$_2$O ice mixture at 10 K \citep{Wu2002}. These experimental studies underline the important role of vinyl alcohol as a key species in the C$_2$H$_2$:H$_2$O chemistry linking to the formation of ketene, acetaldehyde, and ethanol on grain surfaces. Energetic processing of C$_2$H$_2$ has also been proposed to form larger hydrocarbons; proton radiolysis of pure C$_2$H$_2$ leads to the formation of polyynes (sp-hybridized bond) under astronomically relevant conditions \citep{Compagnini2009}. 

In molecular clouds, OH-radicals are expected to be abundantly produced through barrierless atomic reactions such as O + H $\rightarrow$ OH and O$_2$ + 2H $\rightarrow$ 2OH on dust grains and considered as indispensable precursor of interstellar H$_2$O ice \citep{Cuppen2007}. Therefore, the direct OH-radical attachment to the available C$_2$H$_2$ ice offers an alternative reaction route forming COMs without the need for suprathermal O-atoms originating from energetic processing (i.e., dissociation reactions). Very recently, the surface chemistry of C$_2$H$_2$ with "non-energetic" OH-radicals and H-atoms has been experimentally investigated under translucent cloud conditions \citep{Chuang2020}. Here, the term "non-energetic" refers to "thermalized" species reacting with interstellar ice and has been conventionally used to contrast with energetic chemical triggers, such as photons and fast ions as well as electrons \citep{Watanabe2008, Linnartz2015, Arumainayagam2019}. It should be noted that non-energetic reactions indisputably involve energy changes. This study validated the formation of several chemically linked C$_2$H$_{\text{n}}$O molecules, such as vinyl alcohol, acetaldehyde, ketene, and ethanol, along with the H$_2$O ice-forming sequence. The continuous impact of energetic particles or cosmic ray-induced UV-photons will also energetically manipulate the surface chemistry of accumulated ice mantles on dust grains. Therefore, the similar interactions between unsaturated hydrocarbons and available H-atoms or OH-radicals released from H$_2$O dissociation are also expected to occur in ice bulks, where C$_2$H$_2$ is preserved in H$_2$O-rich ice, or these interactions take place at the interface region between hydrogenated amorphous carbon (HAC) dust and the H$_2$O ice layer.

In parallel with non-energetic processing in molecular clouds, this work is aimed at investigating the role of energetic processing of unsaturated hydrocarbons, which are deeply buried in or beneath H$_2$O ice, through impacting cosmic rays before the ice thermally desorbs due to protostar heating. The experimental study presented here is motivated by the abundant detection of C$_2$H$_2$ with its fragments C$_2$H and simultaneous detection of chemically linked C$_2$H$_{\text{n}}$O species across various stages of star formation showing distinct COMs composition. More specifically, we investigated the proton (H$^+$) induced chemistry for C$_2$H$_2$:H$_2$O mixed and layered ice analogs in order to mimic the two scenarios: C$_2$H$_2$ preserved in and capped by H$_2$O ice. The evolution of newly formed COMs, is derived as a function of the accumulated energy dose. The experimental details are described in the next section. The results are presented in Section 3 and discussed in Section 4. Section 5 focuses on the astronomical relevance of the new data presented here and gives our conclusions.    

\begin{table*}[]
        \caption{Summary of IR band strength values used for the species analyzed in this work.}             
        \label{Table1}      
        \centering                          
        \begin{tabular}{ccccc}
                \hline
                \hline
                Species & Chemical formula & IR peak position & Band strength value & Reference \\
                & & (cm$^{-1}$) & (cm molecule$^{-1}$) & \\
                \hline
                Acetylene & C$_2$H$_2$ & 3226 & 3.60E-17 & This work \\
                Water & H$_2$O & 3329 & 2.00E-16 & \cite{Allamandola1988} \\
                Ketene & CH$_2$CO & 2133 & 1.20E-17 & \cite{Berg1991} \\
                Acetaldehyde & CH$_3$CHO & 1135 & 4.30E-18 & \cite{Bennett2005b} \\
                Vinyl alcohol & CH$_2$CHOH & 1147 & 3.10E-17 & \cite{Bennett2005b} \\
                Ethanol & CH$_3$CH$_2$OH & 1046 & 1.40E-17 & \cite{Moore1998} \\
                \hline
        \end{tabular}
\end{table*}

\section{Experimental details}

All experiments were performed using the ultra-high vacuum (UHV) apparatus located at the Laboratory for Experimental Astrophysics in Catania. The details of the experimental setup and the latest modifications with the updated calibrations have been described in \cite{Strazzulla2001} and \cite{Baratta2015}. Here, only the relevant information of this laboratory work is present. A sample holder using a KBr window as substrate is mounted on the tip of a closed-cycle helium cryostat and positioned at the center of the UHV chamber. The substrate temperature monitored by a silicon diode with <1 K relative accuracy can be regulated between 17 and 300 K by an Oxford temperature controller equipped with a resistive heater. The base pressure in the main chamber is $\sim$2$\times$10$^{-9}$ mbar at room temperature. The KBr substrate is fixed at an angle of 45$^{\circ}$ for both the infrared (IR) beam and the ion beam path.
The proton (200 keV H$^+$) source used for the energetic processing of the C$_2$H$_2$ ice is comprised of a Danfysik 1080-200 ion implanter installed in a separate vacuum line (with a base pressure of $\sim$1$\times$10$^{-7}$ mbar), which is connected to the UHV chamber by a UHV gate valve. The ion beam is guided to the main chamber through ion optics, and its spot is electrostatically swept to provide a uniform bombardment cross-section with an area of 1 cm$^2$ covering the entire diagnostic sampling region of the IR beam. The ion fluence is  recorded in situ by a current integrator and further converted to an energy dose using the stopping power estimated by the SRIM simulation for each experiment \citep{Ziegler2011} via the equation:
\begin{equation}\label{Eq1}
\text{Energy dose}_{\text{(200 keV)}}=\textit{S}_{\text{(200 keV)}}\times\textit{F$_{(200 keV)}$},
\end{equation}
where \textit{S}$_{(200~\text{keV})}$ is the stopping power for 200 keV H$^+$ in units of eV cm$^2$ per 16u-molecule, and \textit{F$_{(200 keV)}$} is the total number of impinging 200 keV H$^+$ per area in units of protons cm$^{-2}$. 

Gaseous C$_2$H$_2$ (Air Liquide; $\geq$ 99.6\%) and H$_2$O (Sigma-Aldrich Chromasolv Plus) or  H$_2$$^{18}$O vapor (Sigma-Aldrich; $\geq$ 97\%), purified through multiple freeze-pump-thaw cycles, are introduced into the main chamber through an all-metal needle valve. Ice samples are deposited by applying "background deposition," ensuring a rather uniform structure. Two types of ice samples are prepared: "mixed" and "layered" samples. For the deposition of the mixed ice sample, gases of C$_2$H$_2$ and H$_2$O are pre-mixed in a gas-mixing chamber and then condensed on the pre-cooled KBr substrate at 17 K. For the layered ice sample, a sequential deposition of C$_2$H$_2$ gas and H$_2$O vapor is applied and repeated multiple times to reach the same H$_2$O ice thickness as that of the mixed samples. The applied geometry of layered ice, consisting of depositing C$_2$H$_2$ first and followed by H$_2$O, is not only intended to mimic the interstellar ice mantle, where H$_2$O ice is accumulated on top of HAC dust, but also to amplify the signal of ongoing interface reactions. The growth of the ice sample is strictly limited to the front side, which is realized by a special design of the sample holder (see \citealt{Sicilia2012} for details). The ice thicknesses are monitored in situ
 by He-Ne laser interference measurements following the procedure described in \cite{Baratta1998} and \cite{Urso2016} and in the range of 0.54-0.71 $\mu$m, which is below the penetration depth of the impacting 200 keV H$^+$.

The ice sample is monitored using Fourier transform infrared spectroscopy (FTIR) in a range from 400 to 7500 cm$^{-1}$ with 1 cm$^{-1}$ resolution. The IR absorption area is obtained by Gaussian fitting with one standard deviation used as an error bar. This estimation does not account for uncertainties originating from the baseline subtraction procedure. The column density of parent and product species are further derived by the modified Beer-Lambert law converting IR absorbance area (i.e., optical depth area multiplied by the converting factor of ln 10) to absolute abundance. The applied absorption band strength value of C$_2$H$_2$ has been directly measured on the same experimental setup for C$_2$H$_2$, while other species' values are taken from the literature (see Table \ref{Table1}). The ratio of the predeposited C$_2$H$_2$:H$_2$O (C$_2$H$_2$:H$_2$$^{18}$O) ice mixture is 0.4:1 (0.2:1), where H$_2$O concentration is always overabundant compared to C$_2$H$_2$. The product's abundance ratio of CH$_2$C$^{18}$O over C$^{18}$O as a function of H$^+$ fluence is assumed to be similar to the abundance ratio of CH$_2$CO over CO in a regular H$^+$ radiolysis experiment, where ketene and carbon monoxide cannot be deconvoluted. We note that a recently published absorption band strength for acetaldehyde (see \citealt{Hudson2020}) shows a non-negligible inconsistency for the CH$_3$ deformation mode ($\nu_7$; $\sim$1350 cm$^{-1}$) of acetaldehyde among the available literature values. However, the newly reported band strength, namely 5.3$\times$10$^{-18}$ cm molecule$^{-1}$ for the CH wagging mode ($\nu_8$; $\sim$1135 cm$^{-1}$) of acetaldehyde is close to the value of 4.3$\times$10$^{-18}$ cm molecule$^{-1}$ that is used in this work. The latter value is taken from \cite{Bennett2005b}. In order to minimize the relative uncertainty in the absolute abundances, the band strengths of acetaldehyde and vinyl alcohol are both taken from the theoretical value reported in that work. The utilized absorption peaks of species studied in this work and their corresponding absorption band strength are summarized in Table \ref{Table1}. The obtained column density can easily be recalibrated as soon as more precise values become available.

\section{Results}
\subsection{Product formation of 200 keV H$^+$ radiolysis of C$_2$H$_2$:H$_2$O mixed and layered ices}

\begin{figure*}[]
        \begin{center}
                \includegraphics[width=\textwidth]{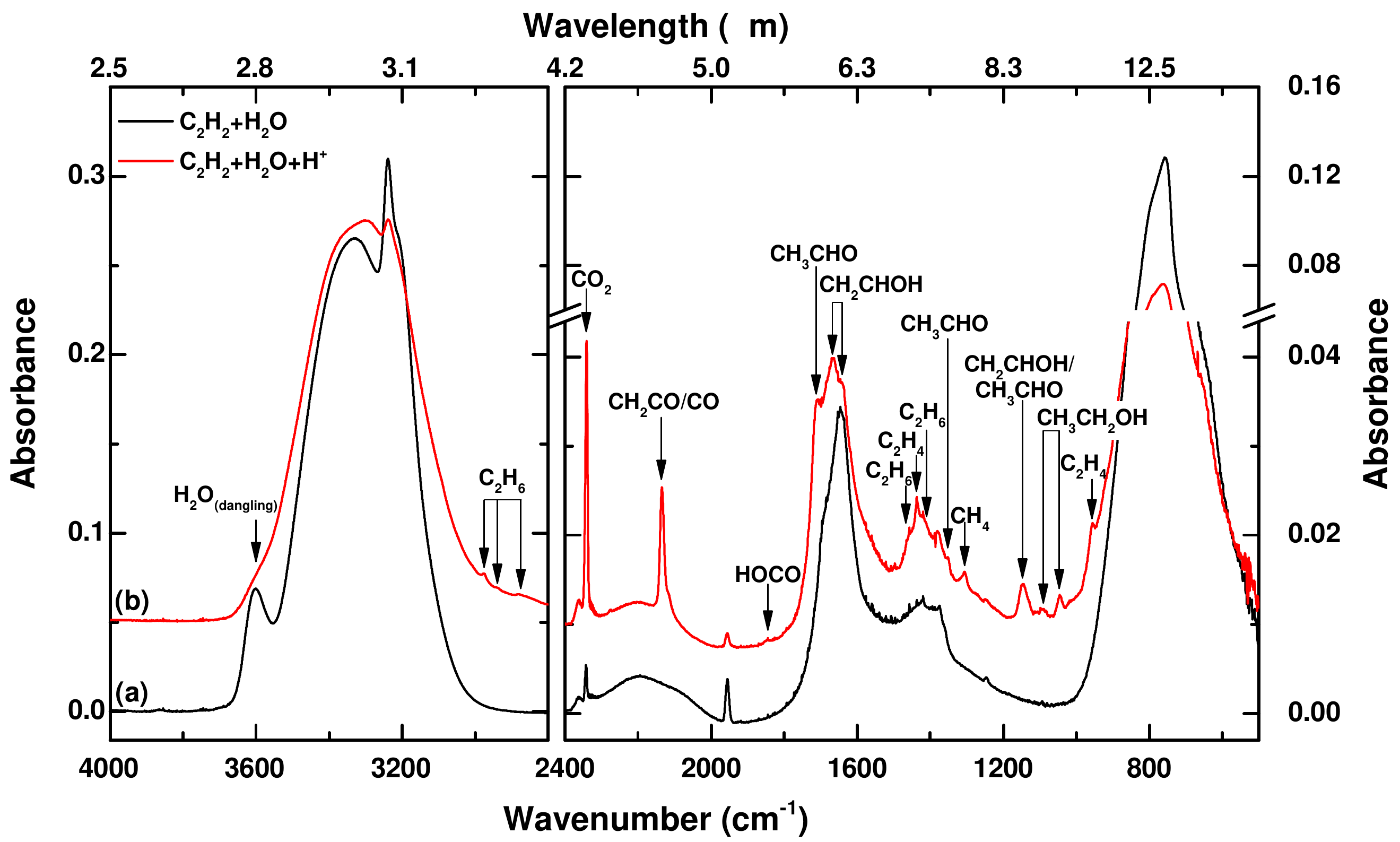}
                \caption{IR spectra obtained after (a, black) deposition of C$_2$H$_2$:H$_2$O mixed ice and (b, red) following ion radiolysis by 200 keV H$^+$ at 17 K. The initial deposition ratio of C$_2$H$_2$:H$_2$O mixed ice was 0.4:1, and the applied H$^+$ fluence was 6.0$\times$10$^{14}$ protons cm$^{-2}$. IR spectra are offset for clarity.}
                \label{Fig1}
        \end{center}
\end{figure*}

Figure 1 presents the IR absorption spectra obtained after (a) deposition of C$_2$H$_2$:H$_2$O mixed ice in a ratio of 0.4:1 and (b) 200 keV H$^+$ radiolysis (i.e., C$_2$H$_2$:H$_2$O + H$^+$) with an H$^+$ fluence of 6.0$\times$10$^{14}$ protons cm$^{-2}$ (i.e., $\sim$ 16.4 eV 16u-molecule$^{-1}$) at 17 K. In the IR spectrum (a) of Fig. 1, the initial C$_2$H$_2$ ice is visible at 758 cm$^{-1}$ ($\nu_5$; CH bending), 1374-1422 cm$^{-1}$ ($\nu_4$+$\nu_5$; combination), and 3239 cm$^{-1}$ ($\nu_3$; CH stretching) \citep{Knez2012, Hudson2014}. The water ice shows up through four broad features at $\sim$742 cm$^{-1}$ (libration), $\sim$1646 cm$^{-1}$ ($\nu_2$; OH bending), $\sim$3329 cm$^{-1}$ ($\nu_1$, $\nu_3$; OH stretching), and $\sim$3600 cm$^{-1}$ ($\nu_R$; OH dangling) \citep{Gerakines1995}. The presence of the H$_2$O dangling band implies a relatively high porosity of the deposited ice mixture. The relatively weak signals at 1955 and $\sim$1230 cm$^{-1}$ originate most likely from low concentration impurities (e.g., C$_\text{n}$H$_\text{m}$) in the acetylene precursor gas that was used. The acetone contamination, which is commonly seen in the acetylene gas bottle, is negligible; the ratio of \textit{N}(CH$_3$COCH$_3$) over deposited \textit{N}(C$_2$H$_2$) is below 0.8\%. In spectrum (b) of Fig. 1, the peak intensity of signals due to C$_2$H$_2$ and H$_2$O significantly decrease after H$^+$ radiolysis. Moreover, the depletion of OH-dangling band is clearly observed suggesting the collapse of the porous structure of the ice bulk upon protons impact \citep{Palumbo2006}. Several new IR features show up; C$_2$H$_4$ can be identified by its vibrational bands at 1437 cm$^{-1}$ ($\nu_{12}$; CH$_2$ scissoring) and 956 cm$^{-1}$ ($\nu_7$; CH$_2$ wagging), and C$_2$H$_6$ exhibits its characteristic features at 2975 cm$^{-1}$ ($\nu_7$; CH asym. stretching), 2939 cm$^{-1}$ ($\nu_5$; CH sym. stretching), 2879 cm$^{-1}$ ($\nu_8$+$\nu_{11}$; CH$_3$ asym. deformation), 1457 cm$^{-1}$ ($\nu_{11}$; CH$_3$ asym. deformation), and 1420 cm$^{-1}$ ($\nu_6$; CH$_3$ sym. deformation) \citep{Shimanouchi1972, Bennett2006, Kaiser2014}. This finding is fully in line with the solid-state hydrogenation scheme of C$_2$H$_2$ $\rightarrow$ C$_2$H$_4$ $\rightarrow$ C$_2$H$_6$, proposed in the literature \citep{Tielens1992, Hiraoka1999, Hiraoka2000, Kobayashi2017, Abplanalp2020, Chuang2020}. The absence of C$_2$H$_3$ and C$_2$H$_5$ implies a low concentration of intermediate radicals, which have been suggested to associate with H-atoms following H$_2$O dissociation. Radiolysis products with a single C-atom are also found in the IR spectrum (b). This is consistent with the previous radiolysis studies by energetic particles, showing efficient and complete dissociation reactions of initial hydrocarbon species \citep{Kaiser2002}. The IR feature at 1305 cm$^{-1}$ originates from the CH bending mode ($\nu_4$) of CH$_4$. The methane C-H stretching mode at 3010 cm$^{-1}$ ($\nu_3$) is blended by the broad H$_2$O band centered at 3329 cm$^{-1}$ \citep{Mulas1998}. CO, CO$_2$, and probably HOCO also shows up through bands at 2135 cm$^{-1}$ (CO stretching), 2341 cm$^{-1}$ ($\nu_3$; O=C=O stretching), and 1845 cm$^{-1}$  ($\nu_2$; C=O stretching), respectively \citep{Milligan1971, Ryazantsev2017}.

Aside from the above assignments of simple hydrocarbons and carbon-bearing molecules, the IR spectrum (b) of Fig. 1 also shows the formation of oxygen-bearing COMs in the C$_2$H$_2$:H$_2$O + H$^+$ experiment. For example, vinyl alcohol is identified through its CO stretching ($\nu_9$), C=C stretching ($\nu_5$), and Fermi resonance with the CH$_2$ wagging overtone (2$\nu_{13}$) at 1147, 1640, and 1665 cm$^{-1}$, respectively \citep{Hawkins1983, Rodler1984, Koga1991}. The latter two peaks severely overlap with the initial broad H$_2$O-dominant feature at $\sim$1646 cm$^{-1}$ and the newly formed peak at 1710 cm$^{-1}$ that is assigned to the C=O stretching mode ($\nu_4$) of acetaldehyde. The full identification of acetaldehyde is supported by the simultaneous detection of the CH$_3$ deformation mode ($\nu_7$) and C-C stretching mode ($\nu_8$) at 1350 and 1135 cm$^{-1}$, respectively \citep{Hollenstein1971, Hawkins1983, Bennett2005b, TerwisschavanScheltinga2018, Hudson2020}. The IR features around 1140 cm$^{-1}$ correspond to the CO stretching mode of vinyl alcohol and CC stretching mode of acetaldehyde as has been confirmed in a previous isotope-labeled study using C$_2$H$_2$ + $^{(18)}$OH + H (see Fig. 2 in \citealt{Chuang2020}). In the present work, therefore, a spectral deconvolution is applied using Gaussian fitting. 
As reported in \cite{Chuang2020}, ethylene oxide, with the highest internal energy among the three C$_2$H$_4$O isomers, is absent in the present work. This result also confirms the proposed formation channel of ethylene oxide, which can be only produced in C$_2$H$_4$ containing ice reacting with O-atoms under ISM-like conditions \citep{Bennett2005b, Ward2011, Bergner2019}. The hydrogen saturated product ethanol is identified by its non-overlapped IR peaks at 1046 and 1088 cm$^{-1}$ that correspond to the CO stretching modes ($\nu_{11}$) and CH$_3$ rocking mode ($\nu_{10}$), respectively \citep{Barnes1970, Mikawa1971, Boudin1998}. The IR feature of ketene shown at 2135 cm$^{-1}$ overlaps with the absorption peak of CO \citep{Hudson2013}. The above identifications of O-bearing COMs are fully consistent with the assignment in the previous non-energetic study of OH-radical and H-atom addition reactions to C$_2$H$_2$ at 10 K \citep{Chuang2020}. 

\begin{figure*}[]
        \begin{center}
                \includegraphics[width=\textwidth]{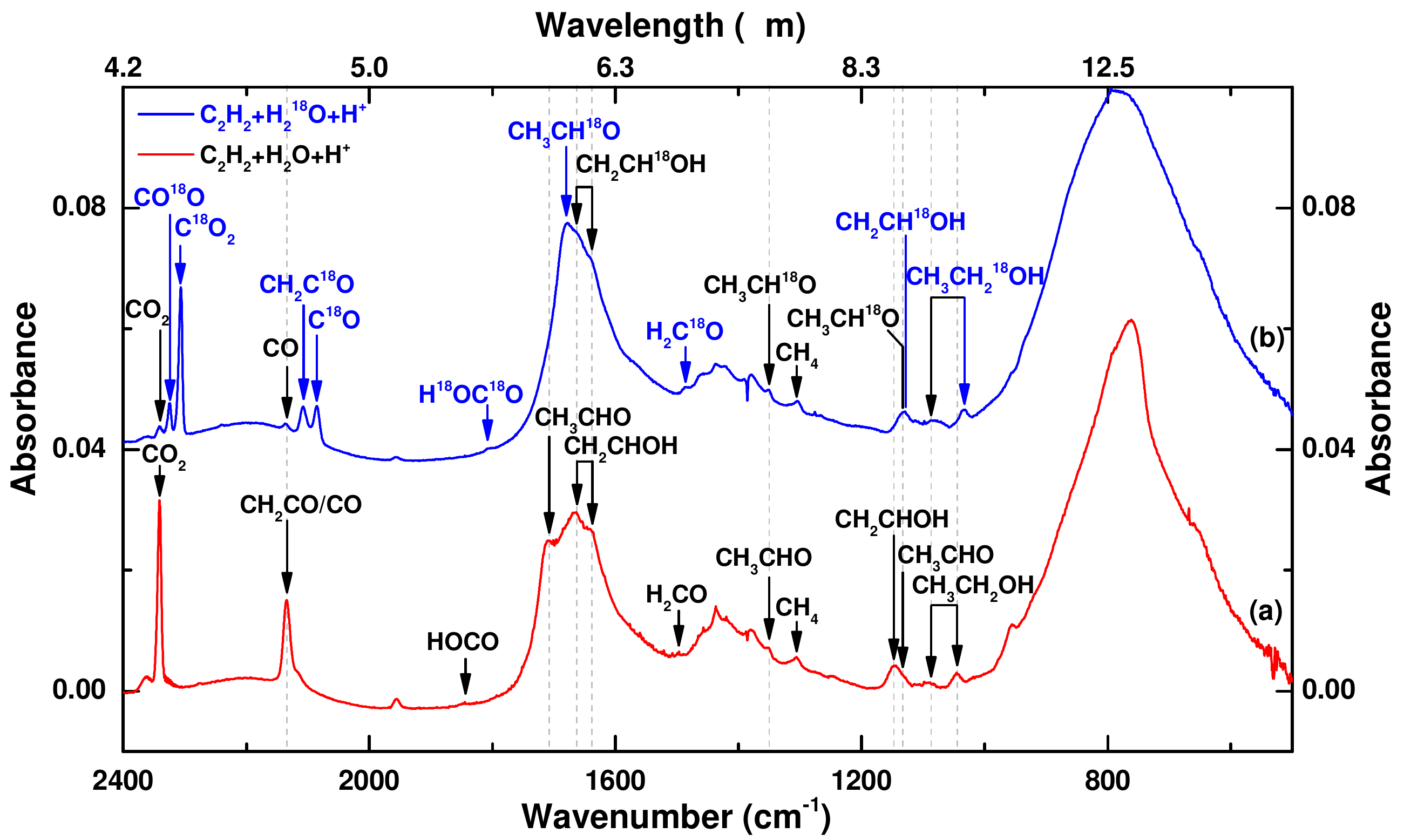}
                \caption{IR spectra obtained after ion bombardment of (a) C$_2$H$_2$:H$_2$O and (b) C$_2$H$_2$:H$_2^{18}$O mixed ices by 200 keV H$^+$ at 17 K. The applied deposition ratios were 0.4:1 and 0.2:1 for C$_2$H$_2$:H$_2$O and C$_2$H$_2$:H$_2^{18}$O mixed ices, respectively. The applied H$^+$ fluence for each experiment is 6.0$\times$10$^{14}$ protons cm$^{-2}$ ($\sim$16.4 eV 16u-molecule$^{-1}$). The blue arrows indicate features that are shifted because of $^{18}$O and the dashed lines reflect the absorption features of COMs with $^{16}$O. IR spectra are offset for clarity.}
                \label{Fig2}
        \end{center}
\end{figure*}

The identifications of these COMs are further secured by utilizing isotope-labeled water ice (i.e., H$_2$$^{18}$O). The comparison between the IR spectra obtained after 200 keV radiolysis of (a) C$_2$H$_2$:H$_2$O and (b) C$_2$H$_2$:H$_2$$^{18}$O ice mixtures under very similar experimental conditions at 17 K is shown in Fig. 2. In addition to simple products, such as C$^{18}$O$_2$ (CO$^{18}$O), C$^{18}$O, H$^{18}$OC$^{18}$O, and possibly H$_2$C$^{18}$O, there are also newly formed oxygen-bearing COMs exhibiting redshifted vibrational modes involving $^{18}$O-atoms. For example, the absorption features of CH$_2$C$^{18}$O and C$^{18}$O are located at 2108 and 2085 cm$^{-1}$, respectively \citep{Hudson2013, Maity2014, Bergner2019}. The C=$^{18}$O stretching mode of acetaldehyde shifts from 1710 to 1678 cm$^{-1}$, while its other vibrational transitions (not involving $^{18}$O-atoms), such as the CH$_3$ deformation mode ($\nu_7$) and C-C stretching mode ($\nu_8$), stay fixed at 1352 and 1135 cm$^{-1}$ \citep{Hawkins1983, Rodler1984}. Vinyl alcohol is also confirmed by detecting its non-shifted vibrational modes, such as CC stretching ($\nu_5$) and Fermi resonance with the CH$_2$ wagging overtone (2$\nu_{13}$) at 1678 and 1640 cm$^{-1}$, while the C$^{18}$O stretching mode ($\nu_9$) shifts from 1147 to 1131 cm$^{-1}$ \citep{Hawkins1983, Rodler1984}. The identification of ethanol is supported by detecting the CH$_3$ rocking mode ($\nu_{10}$) at 1088 cm$^{-1}$ and C$^{18}$O stretching modes ($\nu_{11}$), which redshift from 1046 to 1035 cm$^{-1}$ \citep{Bergner2019}. These IR spectral (non-)shifts strongly support the previous assignments of complex molecules, detected in the 200 keV H$^+$ radiolysis of C$_2$H$_2$:H$_2$O mixed ice at 17 K. 

\begin{figure*}[]
        \begin{center}
                \includegraphics[width=\textwidth]{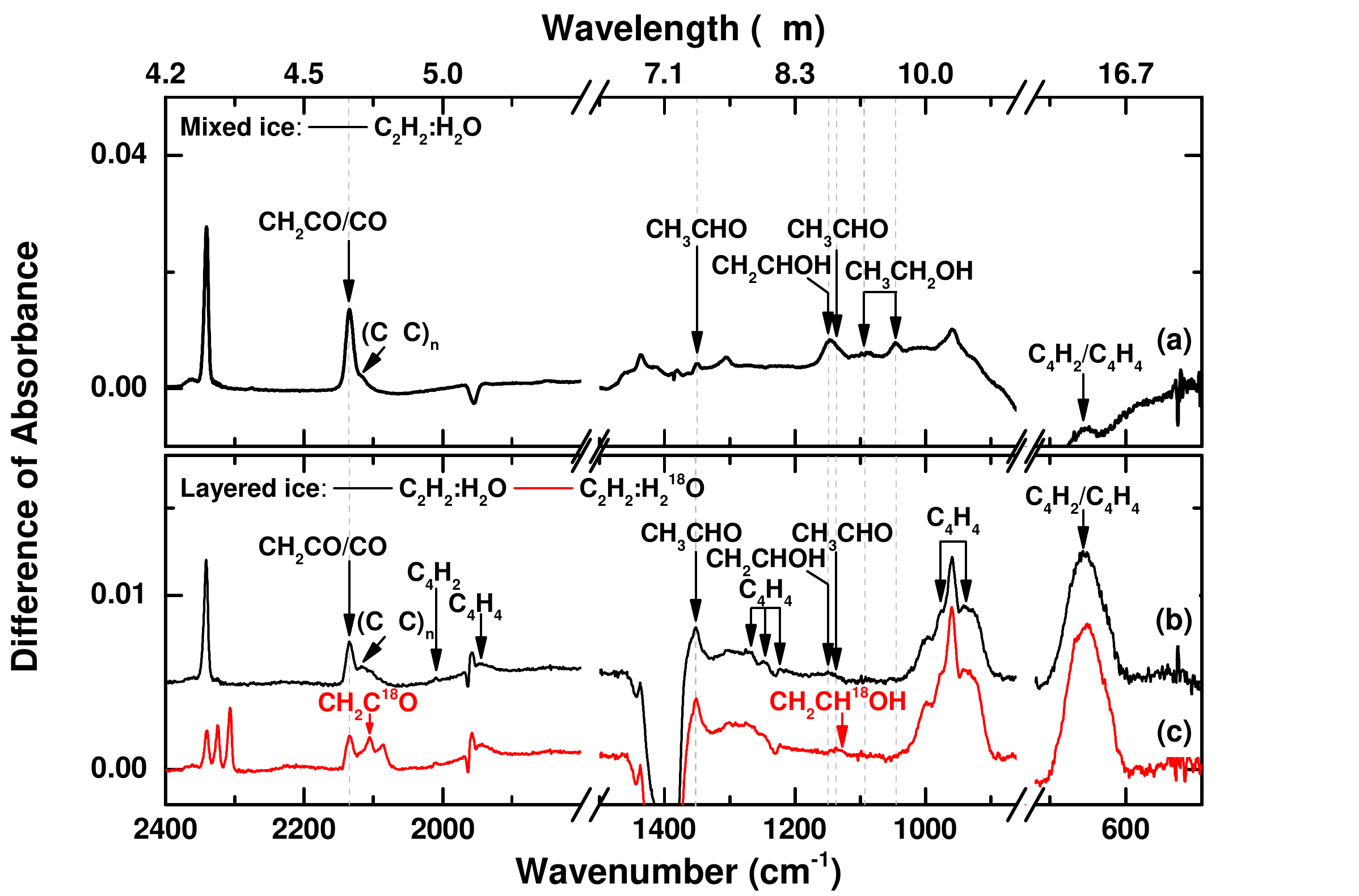}
                \caption{IR difference spectra obtained before and after 200 keV H$^+$ radiolysis of (a) C$_2$H$_2$:H$_2$O mixed ice along with (b) C$_2$H$_2$:H$_2$O and (c) C$_2$H$_2$:H$_2^{18}$O layered ices for the same H$^+$ fluence of 6.0$\times$10$^{14}$ protons cm$^{-2}$ (i.e., $\sim$16.4 eV 16u-molecule$^{-1}$) at 17 K. The black arrows followed by dashed lines indicate the absorption peaks of regular COMs with $^{16}$O-atoms, and red arrows indicate the absorption peaks of COMs  with $^{18}$O-atoms. IR spectra are offset for clarity.}
                \label{Fig3}
        \end{center}
\end{figure*}

In order to directly compare the radiolysis products of two sample geometries (i.e., mixed and layered ices), IR difference spectra are utilized to highlight the newly formed absorption features obtained after proton impact by subtracting parent molecular IR peaks. Figure 3 presents the difference spectra for (a) C$_2$H$_2$:H$_2$O mixed ice, (b) C$_2$H$_2$:H$_2$O layered ice, and (c) C$_2$H$_2$:H$_2^{18}$O layered ice, obtained at 17 K before and after 200 keV H$^+$ radiolysis for a fluence of 6.0$\times$10$^{14}$ protons  cm$^{-2}$ (i.e., $\sim$16.4 eV 16u-molecule$^{-1}$). As mentioned previously, the studied hydrocarbon chemistry with H-atoms or OH-radicals is expected to take place in two relevant interstellar ice mantle regions; one is in the ice bulks where C$_2$H$_2$ is fully mixed with H$_2$O-rich ice, and another is at the contact surface between H$_2$O ice and HAC. In Fig. 3, depleted peaks of C$_2$H$_2$ and H$_2$O are intentionally omitted (except for the one at $\sim$1400 cm$^{-1}$) for clarity. In spectrum (a), the newly formed hydrocarbons (e.g., C$_2$H$_4$ and C$_2$H$_6$) and COMs (i.e., ketene, vinyl alcohol, acetaldehyde, and ethanol) are seen as positive peaks. Besides the R-(C$\equiv$C)-R peak at 2117 cm$^{-1}$, a broad feature at $\sim$656 cm$^{-1}$ is tentatively revealed after subtracting the libration peak of H$_2$O, hinting at the formation of the simplest cumulene (C$_4$H$_4$) and polyyne (C$_4$H$_2$). These two products have been reported in the literature investigating energetic processing of pure C$_2$H$_2$ ice \citep{Compagnini2009, Abplanalp2020, Lo2020, Pereira2020}. In spectrum (b), the absorption peak at 656 cm$^{-1}$ is significantly enhanced compared to spectrum (a). Moreover, several peaks originating from C$_4$H$_4$ are observed at 941 cm$^{-1}$ ($\nu_{15}$), 975 cm$^{-1}$ ($\nu_{14}$), 1223/1248/1276 cm$^{-1}$ (2$\nu_{17}$/$\nu_{11}$+$\nu_{17}$/2$\nu_{11}$), $\sim$1943 cm$^{-1}$ (2$\nu_{15}$/2$\nu_{14}$), and broad 3282 cm$^{-1}$ ($\nu_1$; not shown in Fig. \ref{Fig3}) \citep{Torneng1980, Kim2009}. C$_4$H$_2$ can also be identified by its $\nu_5$ vibration at 2010 cm$^{-1}$, while its two strong absorption peaks located at $\sim$652 cm$^{-1}$ ($\nu_8$/$\nu_6$) and $\sim$3277 cm$^{-1}$  ($\nu_4$) cannot be discriminated from eventual C$_4$H$_4$ bands \citep{Khanna1988, Zhou2009}. Additionally, the spectral assignments of C$_4$H$_4$ and C$_4$H$_2$ are secured by observing the non-shifted peaks in the 200 keV H$^+$ radiolysis of C$_2$H$_2$:H$_2^{18}$O layered ice experiment in the spectrum (c) of Fig. 3. The enhancement of C$_4$H$_2$ and C$_4$H$_4$, which is observed in spectra (b) and (c) representing the pure C$_2$H$_2$ solid-state reactions in layered experiments, implies direct interactions between neighboring C$_2$H$_2$ molecules and is consistent with the non-detection of C$_4$H$_\text{n}$ (n=2 and 4) reported in the experiments with mixed C$_2$H$_2$:H$_2$O ice \citep{Moore1998}. This bottom-up chemistry from the simplest alkyne has been proposed as a solid-state pathway enriching longer (hydro-)carbon chains and even (poly-)aromatic hydrocarbons in the ISM \citep{Necula2000, Tielens2013}.

In addition to the formation of C$_2$H$_2$ derivatives, such as polymers, cumulenes, and polyynes, spectrum (b) in Fig. 3 shows other products as visible from spectrum (a), which only can be explained through interface reactions between the C$_2$H$_2$ and H$_2$O icy layers. The tentative identification of the IR characteristic features of O-bearing COMs is guided by the mixed ice experiments (i.e., spectrum (a) of Fig. 3). In spectrum (b), the IR absorption features of acetaldehyde, vinyl alcohol, and ketene are found at 1352 and 1134, 1149, as well as 2135 cm$^{-1}$, respectively. Other characteristic IR peaks of these COMs are overlapped with the severely depleted features of C$_2$H$_2$ and H$_2$O in the omitted regions. The isotope-labeled COMs (e.g., CH$_2$C$^{18}$O, CH$_2$CH$^{18}$OH) in spectrum (c) provide extra evidence for the correct assignments through the involved red-shifted peaks upon processing of the C$_2$H$_2$:H$_2$$^{18}$O layered ice. In general, the intensity of the observed COM peaks in the spectra (b) and (c) is smaller than that shown in the spectrum (a) as would  intuitively be expected as well. The difference in product yield is due to the limited availability of C$_2$H$_2$ for possible interactions with H$_2$O dissociated fragments; both reactants can only meet at the interface of the layered ice. 

The present experimental results confirm that in the solid state, the simplest alkyne (i.e., C$_2$H$_2$) reacts with adjacent C$_2$H$_2$ fragments (e.g., C$_2$H, CH,  and C$_2$) and H-atoms, forming larger hydrocarbons and actively participates in the formation of O-bearing COMs through interactions with H$_2$O radiolysis products, such as OH-radials and H-atoms. The quantitative analysis of the newly formed COMs and the possible reaction network are discussed in the next sections.

\subsection{Kinetics of H$^+$ radiolysis of C$_2$H$_2$:H$_2$O mixed ice}

The assigned IR features have been recorded as a function of H$^+$ fluence. As mentioned in Section 2, the integrated IR band intensities can be converted to column densities for known absorption band strengths. The resulting abundance evolution, both of the parent species and the newly formed COMs is then sufficient to reveal the reaction kinetics induced by the H$^+$ radiolysis.
 
In Fig. 4, the abundance evolution is shown for the parent C$_2$H$_2$ (i.e., \textit{N}(C$_2$H$_2$)) in the left panel and for the newly formed vinyl alcohol, acetaldehyde, ketene, and ethanol in the right panel upon H$^+$ radiolysis of the C$_2$H$_2$:H$_2$O mixed ice for a H$^+$ fluence up to 6.0$\times$10$^{14}$ protons cm$^{-2}$ (i.e., $\sim$16.4 eV 16u-molecule$^{-1}$). The resulting abundances are normalized to the initial C$_2$H$_2$ amount, namely \textit{N}$_0$(C$_2$H$_2$). The value of \textit{N}(C$_2$H$_2$)/\textit{N}$_0$(C$_2$H$_2$) is rapidly decreasing at the beginning of the H$^+$ radiolysis experiment, and the depletion behavior starts slowing down at a H$^+$ fluence of $\sim$1.5$\times$10$^{14}$ protons cm$^{-2}$ (i.e., $\sim$4.1 eV 16u-molecule$^{-1}$). At the end of the H$^+$ radiolysis experiment, about 66\% of the initial C$_2$H$_2$ is consumed. A very similar C$_2$H$_2$ depleting ratio (i.e., $\sim$65\%) is observed in the isotope-labeled experiment C$_2$H$_2$:H$_2$$^{18}$O + H$^+$. The time-resolved evolution of the C$_2$H$_2$ abundance is further fitted by a single exponential equation (see Eq. 3 in \citealt{Garozzo2011}):
\begin{equation}\label{Eq2}
\frac {\textit{N}(\text{C$_2$H$_2$})}{N_0(\text{C$_{2}$H$_{2}$})}=(1-\alpha)\cdot\text{exp}(-\sigma\cdot\textit{F})+\alpha,
\end{equation}
where $\alpha$ is the saturation value (unitless), \textit{F} is the H$^+$ fluence in protons cm$^{-2}$, and $\sigma$ is the effective reaction cross-section in cm$^2$ proton$^{-1}$. The derived destruction cross-section is $\sim$(1.9$\pm$0.5)$\times$10$^{-15}$ cm$^2$ per H$^+$ (200 keV), which is equal to $\sim$0.07$\pm$0.02 16u-molecules eV$^{-1}$. 

\begin{figure}[t]
        \begin{center}
                \includegraphics[width=90mm]{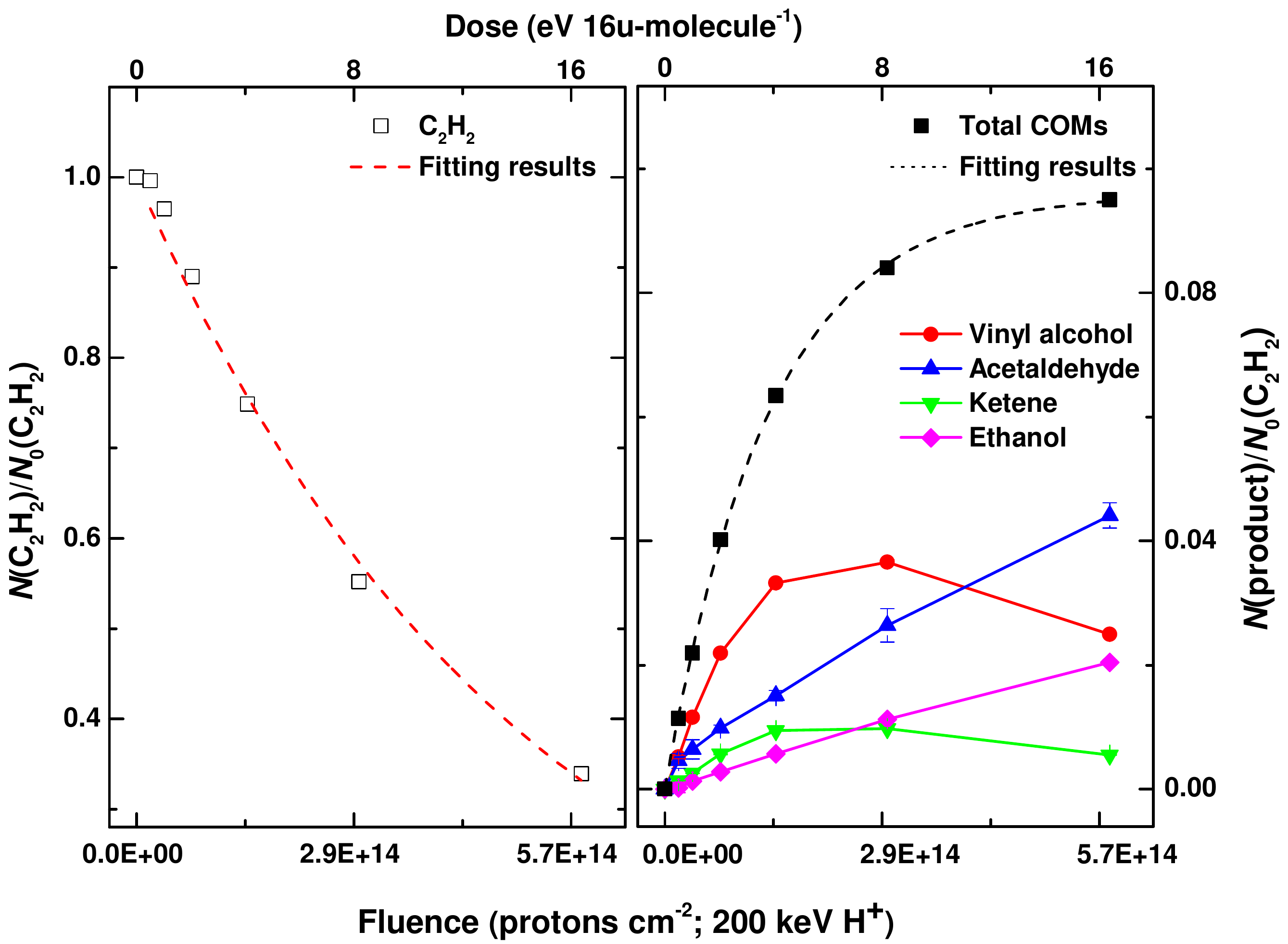}
                \caption{Abundance evolution of the parent C$_2$H$_2$ species (left panel) and the newly formed reaction products (right panel) in the 200 keV H$^+$ radiolysis of C$_2$H$_2$:H$_2$O mixed ice over H$^+$ fluence of 6.0$\times$10$^{14}$ protons cm$^{-2}$ (i.e., $\sim$16.4 eV 16u-molecule$^{-1}$). The derived absolute abundances are normalized to the initial C$_2$H$_2$ abundance. The dashed lines present the fitting results, and the solid lines connecting data are only for clarity.}
                \label{Fig4}
        \end{center}
\end{figure}

In the right panel of Fig. 4, the abundance evolution of newly formed products as a function of H$^+$ fluence is shown, illustrating that different species follow different evolutionary tracks. For example, vinyl alcohol is immediately and abundantly formed upon H$^+$ radiolysis, implying that vinyl alcohol is a first-generation product. Later, the increasing \textit{N}(CH$_2$CHOH)/\textit{N}$_0$(C$_2$H$_2$) value noticeably slows down around a fluence of $\sim$1.5$\times$10$^{14}$ protons cm$^{-2}$, which coincides with C$_2$H$_2$ depletion behavior, and reaches its maximum intensity of $\sim$0.037 at a fluence of $\sim$3.0$\times$10$^{14}$ protons cm$^{-2}$. Such a fast-increasing curve that is followed by rapid depletion has also been observed in previous ion-radiolysis studies of C$_2$H$_2$:H$_2$O mixed ice \citep{Hudson2003, Zasimov2020}. A similar kinetic behavior is found in the ketene formation curve, implying a shared chemical history; vinyl alcohol is a precursor of ketene as suggested by \cite{Hudson2013}. This chemical link is additionally supported by observing a constant ratio between ketene and vinyl alcohol (i.e., $\sim$0.25$\pm$0.03) over the entire H$^+$ radiolysis experiment.

Acetaldehyde shows a similarly fast-increasing trend as vinyl alcohol upon start of the H$^+$ radiolysis, representative for a fast enol-keto conversion as observed in previous studies \citep{Abplanalp2016, Chuang2020}. This will be further discussed in section 4. However, the acetaldehyde formation curve does not follow vinyl alcohol (or ketene), which starts depleting at a fluence of $\sim$3.0$\times$10$^{14}$ protons cm$^{-2}$. In contrast, the abundance of acetaldehyde (and also of ethanol) increases continuously with applied H$^+$ fluence. Consequently, the accumulated column density of acetaldehyde at the end of the H$^+$ radiolysis even surpasses the amount of vinyl alcohol. The final column densities of complex products with respect to the initial C$_2$H$_2$ abundance (i.e., \textit{N}(products)/\textit{N}$_0$(C$_2$H$_2$)) are obtained with values of $\sim$0.044, 0.025, 0.020, and 0.006 for acetaldehyde, vinyl alcohol, ethanol, and ketene, respectively. The composition ratios for the reaction products are summarized in Table \ref{Table2}.
 
The cumulative abundance of all COMs is shown as a black curve in the right panel of Fig. 4 and can be fitted by a single exponential equation:
\begin{equation}\label{Eq15}
\frac {\textit{N}(\text{Total COMs})}{N_0(\text{C$_{2}$H$_{2}$})}=\alpha(1-\text{exp}(-\sigma\cdot\textit{F})),
\end{equation}
where $\alpha$ is the saturation value (unitless), \textit{F} is the H$^+$ fluence in protons cm$^{-2}$, and $\sigma$ is the effective reaction cross-section in cm$^2$ proton$^{-1}$. This results in an effective formation cross-section of $\sim$(7.2$\pm$0.1)$\times$10$^{-15}$ cm$^2$ per H$^+$ (i.e., 0.26$\pm$0.01 16u-molecule eV$^{-1}$) for the overall complex products. The pseudo first-order fitting suggests that these products have a common precursor, which acts as a limited reactant in bimolecular reactions. Although the total product abundance nearly reaches a steady-state, the destruction of C$_2$H$_2$ is still going on (i.e., lower reaction cross-section derived) and resulting in smaller species, such as CH$_4$ and CO \citep{Hudson2013, Zasimov2020}. Therefore, the obtained conversion ratio of the overall complex products to depleted C$_2$H$_2$ (i.e., \textit{N}(products)/$\Delta$\textit{N}(C$_2$H$_2$)) is a rather dynamic value depending on the applied H$^+$ fluence (or energy dose) (see Table \ref{Table2}).

\begin{table*}[]
        \caption{Conversion and composition ratios of products in 200 keV H$^+$ radiolysis of C$_2$H$_2$:H$_2$O mixed ice at 17 K.}             
        \centering                          
        \begin{tabular}{ccccccc}
                \hline
                \hline
                H$^+$ Fluence (200 keV) & Dose & $\Delta$\textit{N}(products)/$\Delta$\textit{N}(C$_2$H$_2$) & \multicolumn{4}{c}{ Product's composition ratio } \\
                protons cm$^{-2}$s$^{-1}$ & eV 16u-molecule$^{-1}$ & & vinyl alcohol & ketene & acetaldehyde & ethanol \\
                \hline
                1.88E+13 & 0.5 & 0.84 & 0.12 & 0.40 & 0.46 & 0.02 \\
                3.80E+13 & 1.0 & 0.63 & 0.53 & 0.12 & 0.29 & 0.06 \\
                7.50E+13 & 2.1 & 0.36 & 0.55 & 0.14 & 0.24 & 0.07 \\
                1.50E+14 & 4.1 & 0.25 & 0.52 & 0.15 & 0.24 & 0.09 \\
                3.00E+14 & 8.2 & 0.19 & 0.44 & 0.12 & 0.31 & 0.13 \\
                6.00E+14 & 16.4 & 0.14 & 0.26 & 0.06 & 0.46 & 0.22 \\
                \hline
        \end{tabular}
        \label{Table2}
\end{table*}

\section{Discussion}

\begin{figure}[b!]
        \begin{center}
                \includegraphics[width=85mm]{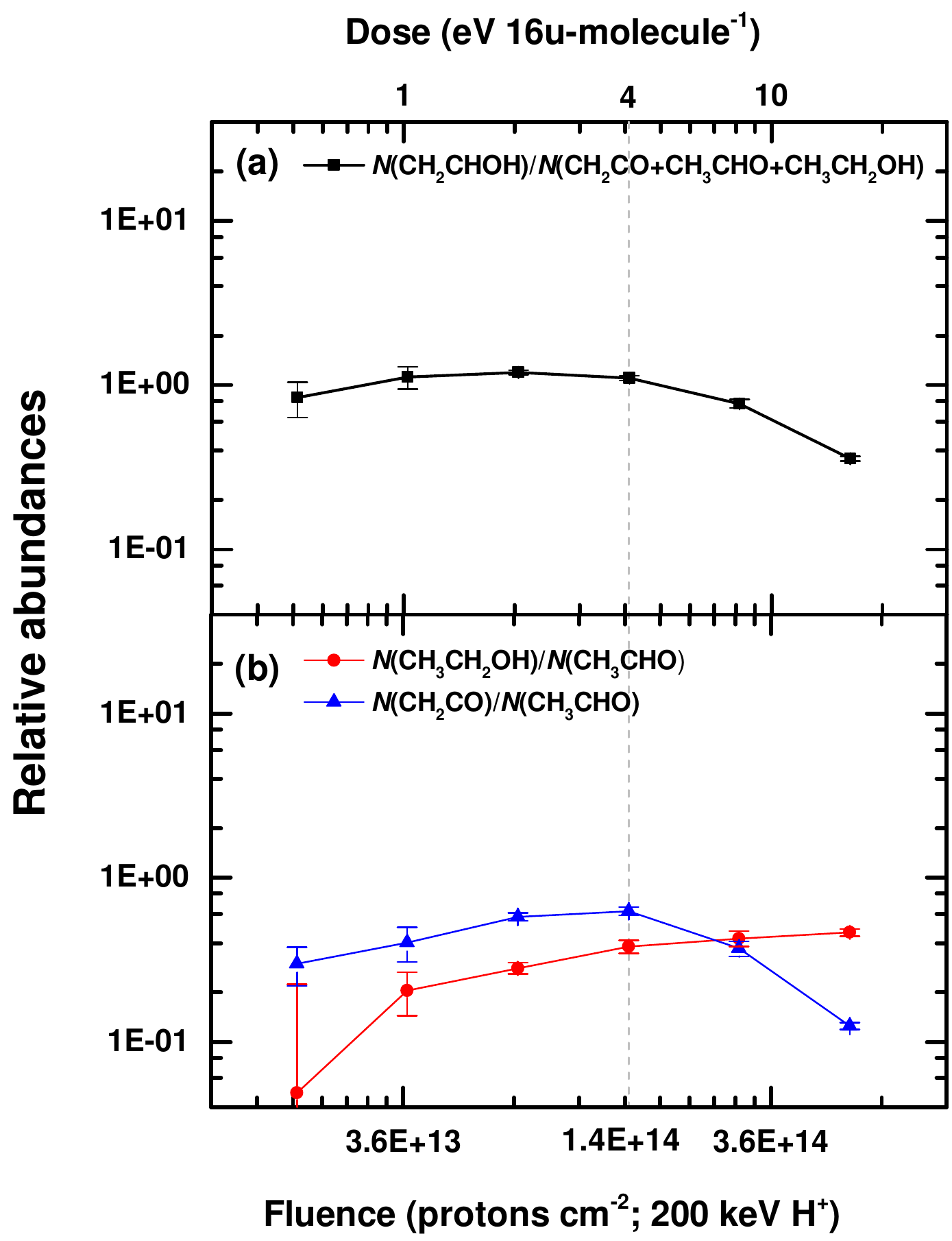}
                \caption{Relative abundances of all observed COMs in the 200 keV H$^+$ radiolysis of C$_2$H$_2$:H$_2$O mixed ice over H$^+$ fluence of 6.0$\times$10$^{14}$ protons cm$^{-2}$ (i.e., $\sim$16.4 eV 16u-molecule$^{-1}$). Upper (a): Relative abundances of vinyl alcohol over the total yields of ketene, acetaldehyde, and ethanol. Bottom (b): Relative abundances of ketene over acetaldehyde and ethanol over acetaldehyde. The dashed line shows the turning point of product's formation behavior. The data points are connected by solid lines for clarity.}
                \label{Fig5}
        \end{center}
\end{figure} 

The H$^+$ radiolysis of ice mixtures on grain surfaces primarily causes species to dissociate and ionize, generating a cascade of (high- and low-energy) electrons. These reactions efficiently take place along the H$^+$ penetrating path until high energy particles (e.g., protons or electrons) run out of their kinetic energy. The ion radiolysis of H$_2$O ice has been intensively studied in the literature (\citealt{Moore2000, Gomis2004, Loeffler2006, Buxton2008, Arumainayagam2019}, and references therein). It has been concluded that the ion-induced excitation of H$_2$O ice (i.e., H$_2$O*) is most likely to fragment into OH-radicals and H-atoms and the ionization of H$_2$O mainly leads to the generation of H$_2$O$^+$ and e$^-$. These ions and electrons can further react with surrounding H$_2$O ice, ultimately enriching the effective abundance of OH-radicals and H-atoms via a series of ion-molecule (e.g., H$_2$O$^+$ + H$_2$O $\rightarrow$ OH + H$_3$O$^+$) or electron-ion (e$^-$ + H$_3$O$^+$ $\rightarrow$ H + H$_2$O) reactions; H$_2$O radiolysis mainly yields OH-radicals and H-atoms as well as some other derivatives, such as H$_2$ and H$_2$O$_2$ \citep{Buxton2008}. Via the insertion of acetylene in this study, the efficient formation of hydrogenated products, such as C$_2$H$_4$ and C$_2$H$_6$, as well as hydroxylated organics (i.e., upon OH-addition), such as vinyl alcohol, is the consequence of these H-atoms and OH-radicals reacting with C$_2$H$_2$. It is important to note that also ion-molecule reactions might occur along with the H$^+$ penetrating path, but these are beyond the scope of this work. A detailed study on solid-state interaction between ions and molecules under interstellar conditions is still required. Here we focus on the reaction network of C$_2$H$_2$ interacting with OH-radicals and H-atoms, induced by H$^+$ radiolysis of H$_2$O ice.

The relative abundances of all observed COMs are shown in Fig. 5, unraveling the possible chemical transformation among vinyl alcohol, acetaldehyde, ketene, and ethanol. In contrast to Fig. 4, the horizontal axis is shown on a logarithmic scale to pinpoint the very initial chemistry upon H$^+$ radiolysis of the C$_2$H$_2$:H$_2$O mixed ice. As discussed in Section 3.2, the acetylene and COM abundance evolution show two distinct formation behaviors, which can be roughly separated at a fluence of $\sim$1.5$\times$10$^{14}$ protons cm$^{-2}$. At the beginning of H$^+$ radiolysis, C$_2$H$_2$ is quickly consumed and converted to vinyl alcohol, and later after passing its maximum yield, the production of acetaldehyde and ethanol takes over. The observed turning point at such fluence is related to the abundance of available C$_2$H$_2$ in the ice mixture; approximately 50\% of the overall C$_2$H$_2$ depletion (i.e., $\Delta$\textit{N}(C$_2$H$_2$)) is reached at a fluence of $\sim$1.5$\times$10$^{14}$ protons cm$^{-2}$.
 
\cite{Hudson2003} identified the formation of vinyl alcohol by observing its IR feature at ~1145 cm$^{-1}$, which was first reported in UV-irradiation of C$_2$H$_2$:H$_2$O ice mixture by \cite{Wu2002}. The authors proposed the chemical link between vinyl alcohol and C$_2$H$_2$ in ion radiolysis of a C$_2$H$_2$:H$_2$O ice mixture at 15 K. The underlying formation steps have recently been investigated in a non-energetic processing experiment through OH-radicals or H-atoms addition reactions with C$_2$H$_2$ on grain surfaces \citep{Chuang2020};

\begin{subequations}
\begin{equation}\label{Eq3a}
\text{C$_2$H$_2$} + \text{OH} \rightarrow \text{C$_2$H$_2$OH}
,\end{equation}
\begin{equation}\label{Eq3b}
\text{C$_2$H$_2$OH} + \text{H} \rightarrow \text{CH$_2$CHOH}
\end{equation}
\end{subequations}
or
\begin{subequations}
        \begin{equation}\label{Eq4a}
        \text{C$_2$H$_2$} + \text{H} \rightarrow \text{C$_2$H$_3$}
        ,\end{equation}
        \begin{equation}\label{Eq4b}
        \text{C$_2$H$_3$} + \text{OH} \rightarrow \text{CH$_2$CHOH}.
        \end{equation}
\end{subequations}

Theoretical calculations show that in the gas phase both association routes require relatively small barriers, such as 5.4–8.0 and 17.99 kJ mol$^{-1}$ for reactions \ref{Eq3a} and \ref{Eq4a}, respectively \citep{Basiuk2004, Miller2004}. The two to three times lower activation energy of reaction \ref{Eq3a} w.r.t. reaction \ref{Eq4a} hints for a preference of vinyl alcohol formation via hydroxylation of C$_2$H$_2$ followed by H-atom addition reactions (i.e., reactions \ref{Eq3a} and \ref{Eq3b}). Moreover, as suggested in theory and laboratory studies, the newly formed vinyl alcohol can immediately be tautomerized (enol$\leftrightarrow$keto) through an "intermolecular" pathway if it is surrounded by other highly catalytic species, such as H$_2$O, C$_2$H$_2$, and acids, forming acetaldehyde \citep{Apeloig1990, Klopman1979, daSilva2010}:
\begin{equation}\label{Eq5}
        \text{CH$_2$CHOH (enol-form)} \leftrightarrow \text{CH$_3$CHO (keto-form)}.
\end{equation}

In Fig. 5 (a), the derived product ratio of vinyl alcohol w.r.t. the sum of the other three COMs studied here is shown;  \textit{N}(CH$_2$CHOH)/\textit{N}(CH$_2$CO+CH$_3$CHO+CH$_3$CH$_2$OH) exhibits a relatively constant value of 1.07$\pm$0.15 before a fluence of $\sim$1.5$\times$10$^{14}$ protons cm$^{-2}$. This implies that the increasing amount of vinyl alcohol is proportional to the total formation yield of ketene, acetaldehyde, and ethanol. Moreover, as shown in Fig. 4, vinyl alcohol is the dominant product at the beginning of the H$^+$ radiolysis. All these results imply that vinyl alcohol acts as a main precursor for the other COMs. The possible conversion reactions are discussed as follows. Besides the isomerization from vinyl alcohol to acetaldehyde, as in reaction (5), the newly formed vinyl alcohol is also expected to be consumed by H$_2$ elimination reactions induced by energetic processes forming ketene \citep{Hudson2013}:
\begin{equation}\label{Eq6}
\text{CH$_2$CHOH}\xrightarrow{\text{+energy}}\text{CH$_2$CO}+\text{2H (or H$_2$),}
\end{equation}
or by hydrogen addition reactions forming ethanol:
\begin{equation}\label{Eq7}
\text{CH$_2$CHOH}\xrightarrow{\text{+2H}}\text{CH$_3$CH$_2$OH}.
\end{equation}  

Reactions \ref{Eq6} and \ref{Eq7} have been individually proposed in studies by \cite{Hudson2013} and \cite{Hudson2003}, respectively. Upon reaching a fluence of $\sim$1.5$\times$10$^{14}$ protons cm$^{-2}$, the relative abundance of vinyl alcohol over the total yields of ketene, acetaldehyde, and ethanol drops by $\sim$67\% from 1.10 to 0.36. The depletion time coincides with that of the slowdown of C$_2$H$_2$ consumption shown in Fig. 4 (a), confirming that the formation of vinyl alcohol is limited by the available C$_2$H$_2$ reacting with OH-radicals. Moreover, the conversion from vinyl alcohol to other COMs (i.e., reactions \ref{Eq5}-\ref{Eq7}) is still active, effectively decreasing the ratio between vinyl alcohol and the other detected COMs.
 
Besides reactions \ref{Eq5}-\ref{Eq7} based on the first-generation product vinyl alcohol, all these newly formed COMs are chemically linked through (de-)hydrogenation. Therefore, the reaction scheme of ketene$\leftrightarrow$acetaldehyde$\leftrightarrow$ethanol has also been considered to play an indispensable role in further manipulating the product ratio via H-atom addition reactions (also see Fig. 6):
\begin{equation}\label{Eq8}
\text{CH$_2$CO}\xrightarrow{\text{+2H}}\text{CH$_3$CHO}
\end{equation} 
and 
\begin{equation}\label{Eq9}
\text{CH$_3$CHO}\xrightarrow{\text{+2H}}\text{CH$_3$CH$_2$OH;}
\end{equation} 
or via H-atom abstraction reactions:
\begin{equation}\label{Eq10}
\text{CH$_3$CHO}\xrightarrow{\text{+2H(-4H)}}\text{CH$_2$CO}
\end{equation} 
and 
\begin{equation}\label{Eq11}
\text{CH$_3$CH$_2$OH}\xrightarrow{\text{+2H(-4H)}}\text{CH$_3$CHO}.
\end{equation} 

The non-energetic interactions between H-atoms and acetaldehyde have been experimentally investigated before under dense cloud conditions to validate the reaction network of ketene$\xleftarrow{\text{+2H}}$acetaldehyde$\xrightarrow{\text{+2H}}$ethanol \citep{Bisschop2007, Chuang2020}. It was found that the effective efficiency of abstraction reactions \ref{Eq10} and \ref{Eq11} is generally lower than that of the addition reactions \ref{Eq8} and \ref{Eq9}. A similar trend has also been found for the interactions between H-atoms and H$_2$CO, along the CO-(de)hydrogenation scheme \citep{Hidaka2007, Chuang2018a}. Therefore, these non-energetic dehydrogenation processes are assumed not to affect the complex product composition ratios substantially.

The relative abundances of ketene, acetaldehyde, and ethanol are shown in Fig. 5 (b). The \textit{N}(CH$_2$CO)/\textit{N}(CH$_3$CHO) is enhanced by a factor of $\sim$2, from 0.30 to 0.63 before an H$^+$ fluence of $\sim$1.5$\times$10$^{14}$ protons cm$^{-2}$. The favored ketene formation route is consistent with an additional formation mechanism (i.e., reaction \ref{Eq6}). Otherwise, the ketene to acetaldehyde ratio should decrease along with the H$^+$ fluence because of the successive conversion from ketene to acetaldehyde (i.e., reaction \ref{Eq8}). An enhancement of the ethanol to acetaldehyde ratio is also found from 0.05 to 0.38 before the H$^+$ fluence of $\sim$1.5$\times$10$^{14}$ protons cm$^{-2}$, supporting that the H-atom abstraction (i.e., reaction \ref{Eq11}) is a less important reaction. However, it is difficult to distinguish between contributions from reactions \ref{Eq7} and \ref{Eq8} because both can increase the \textit{N}(CH$_3$CH$_2$OH)/\textit{N}(CH$_3$CHO) ratio.

Once the C$_2$H$_2$ conversion to vinyl alcohol slows down, which is clearly observed after reaching the H$^+$ fluence of $\sim$1.5$\times$10$^{14}$ protons cm$^{-2}$ in Fig. 4, the ratio of \textit{N}(CH$_2$CO)/\textit{N}(CH$_3$CHO) shows a rapid decrease by $\sim$80\% from 0.63 to 0.13 due to the efficient ketene hydrogenation (i.e., reaction \ref{Eq8}). In contrast, the \textit{N}(CH$_3$CH$_2$OH)/\textit{N}(CH$_3$CHO) ratio keeps increasing, from 0.38 to 0.48, but with a much lower efficiency than before. The relative slow increasing rate is probably due to the shortage of vinyl alcohol or the continuous increase of acetaldehyde through reaction \ref{Eq8}. Since the contribution of vinyl alcohol to the other three products becomes a minor route, the successive hydrogenation scheme of ketene$\rightarrow$acetaldehyde$\rightarrow$ethanol further mediates the hydrogen-content of COMs, favoring species such as acetaldehyde and ethanol.

\begin{figure}[t]
        \begin{center}
                \includegraphics[width=80mm]{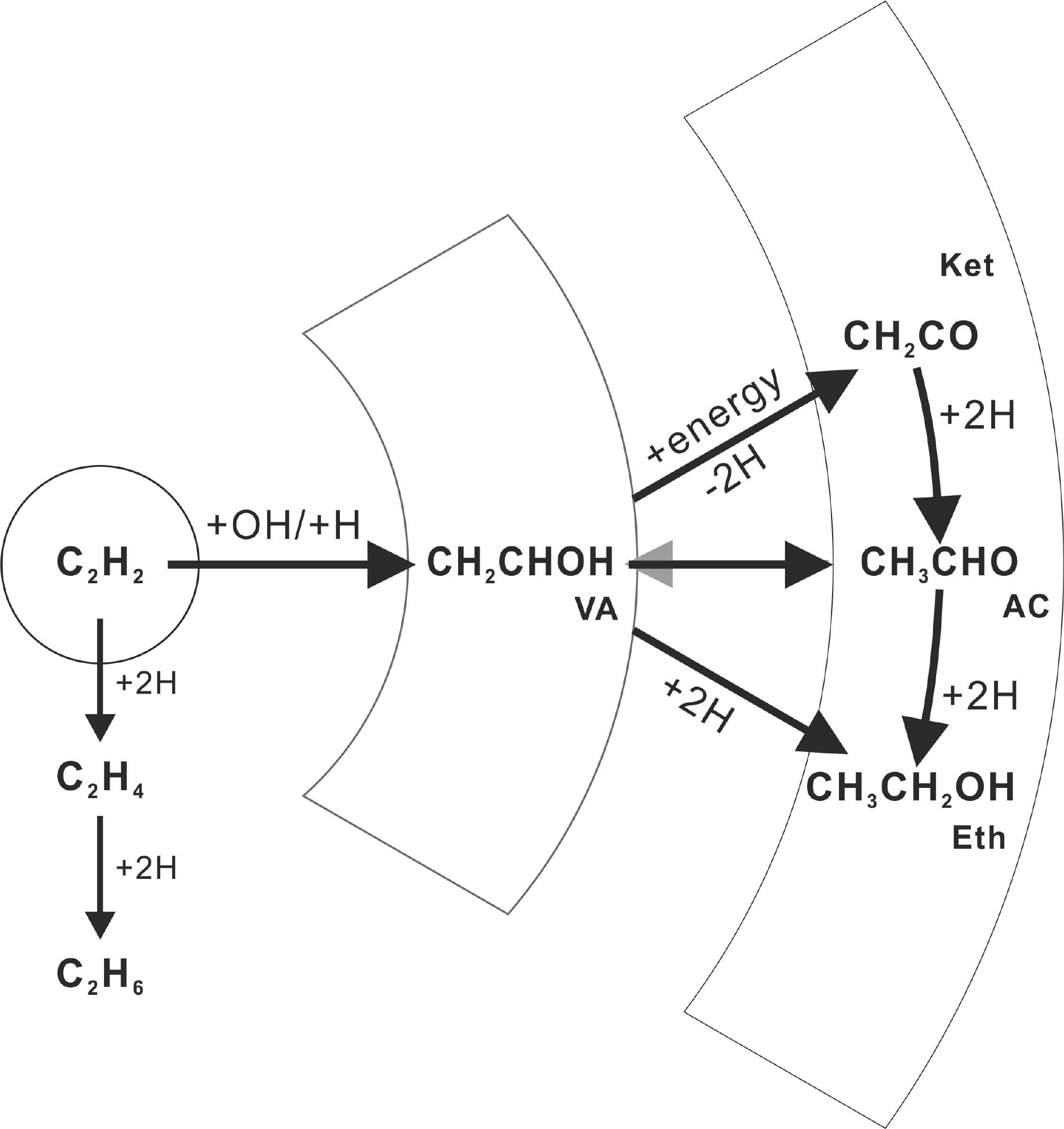}
                \caption{Proposed reaction diagram linking the parent species C$_2$H$_2$ (containing C$\equiv$C bond) and newly formed COMs (containing C=C and C-C bonds) in H$^+$ radiolysis of C$_2$H$_2$:H$_2$O ice mixture. VA: vinyl alcohol; AC: acetaldehyde; Ket: ketene; Eth: ethanol.}
                \label{Fig6}
        \end{center}
\end{figure} 

\section{Astrochemical implication and conclusions}

\begin{figure*}[t]
        \begin{center}
                \includegraphics[width=\textwidth]{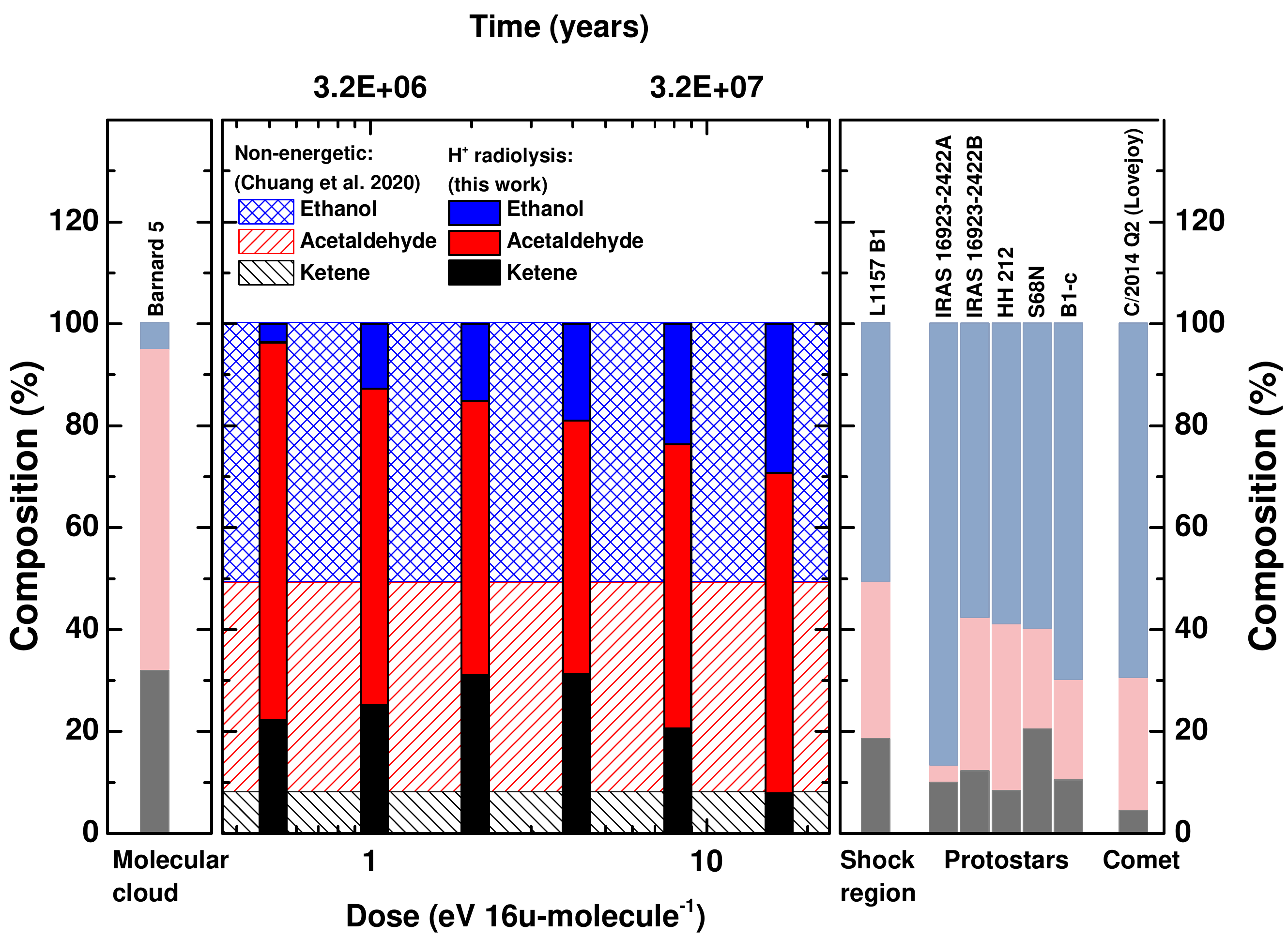}
                \caption{Composition evolution of ethanol, acetaldehyde, and ketene as a function of energy dose in the H$^+$ radiolysis experiment of C$_2$H$_2$:H$_2$O mixed ice at 17 K (middle panel). Here, "time" stands for the approximate years required in dense molecular clouds to reach the same amount of energy dose. The shadow area (in background) reflects the composition ratio in previous non-energetic study obtained from \cite{Chuang2020}. The selected observational data toward dense cloud Barnard 5 (left panel), protostellar shock region L1157 B1, protostar IRAS 16293-2422A, B, HH 212, B1-c, and S68N, as well as comet C/2014 Q2(Lovejoy) (right panel), are obtained from \cite{Taquet2017}, \cite{Lefloch2017}, \cite{Manigand2020}, \cite{Jorgensen2018}, \cite{Lee2019}, \cite{vanGelder2020} and \cite{Biver2019} respectively, standing for an early and later stage in the star-formation process}
                \label{Fig7}
        \end{center}
\end{figure*}
This laboratory study validates a solid reaction network for forming interstellar COMs described by the formula C$_2$H$_{\text{n}}$O, such as vinyl alcohol, acetaldehyde, ketene, and ethanol, through 200 keV H$^+$ radiolysis of interstellar C$_2$H$_2$:H$_2$O ice analogs at 17 K. In this work, we extend the recently studied non-energetic chemistry of C$_2$H$_2$ with OH-radicals and H-atoms to the energetic counterpart triggered by cosmic rays. The initial and newly formed species are  monitored in situ in the solid state by IR spectroscopy. The product identifications are further secured using isotope-labeled precursors (i.e., H$_2$$^{18}$O). The quantitative analysis of the depletion of parent C$_2$H$_2$ and the formation of complex products, which are shown as a function of energy dose, provides information on the involved reaction kinetics. The experimental results are of astronomical relevance to understand the chemistry of unsaturated hydrocarbons deeply buried in H$_2$O ice induced by the continuous cosmic ray impacts along with the evolution of molecular cloud before ice mantle evaporates due to the warming up induced by the protostar.

In the gas phase, C$_2$H$_2$ has been commonly observed in the circumstellar shell of carbon-rich AGB stars such as IRC+10216, together with its photolysis fragment C$_2$H, and it also has been abundantly detected toward interstellar clouds (C$_2$H$_2$/CO = $\sim$10$^{-3}$), massive young stellar objects (YSOs; C$_2$H$_2$/H$_2$O = $\sim$(1-5) $\times$10$^{-2}$), and protoplanetary disks (C$_2$H$_2$/H$_2$O = $\sim$1$\times$10$^{-2}$) \citep{Keady1988, Lacy1989, Cernicharo1999, Lahuis2000, Carr2008}. Although the direct observation of icy C$_2$H$_2$ in the ISM is limited to one observational study by \cite{Knez2008}, a solid-state origin of C$_2$H$_2$ is widely accepted to explain the observed abundance in the gas phase \citep{Sonnentrucker2007}. This is also supported by the observations toward cometary comae, which are regarded as pristine material inherited from interstellar molecular clouds, showing the ratio of C$_2$H$_2$ over H$_2$O is in the range of (0.2-0.9) $\times$10$^{-2}$ \citep{Brooke1996, Mumma2003}. Alternatively, modeling and laboratory studies proposed a top-down formation mechanism of C$_2$H$_2$ through surface erosion of hydrogenated amorphous carbonaceous dust (HAC) or photodissociation of ionized polycyclic aromatic hydrocarbons (PAHs) in harsh UV-photons prevailing regions \citep{Jochims1994, Allain1996, LePage2003, Jager2011, Zhen2014, West2018}.

The investigated solid-state reactions between C$_2$H$_2$ and H$_2$O radiolysis fragments, such as OH-radicals and H-atoms, are shown as a prevalent pathway transforming the simplest alkyne, C$_2$H$_2$, to O-bearing COMs in icy environments where H$_2$O is the major constituent of interstellar ices. In translucent clouds (1$\leq$A$_v$$\leq$5), atomic gases, such as H- and O-atoms, accrete on grain surfaces and associate with available species. For example, H$_2$O ice is mainly formed through the successive H-atom addition reactions to O-atoms:
\begin{equation}\label{Eq12}
\text{O}\xrightarrow{\text{H}}\text{OH}\xrightarrow{\text{H}}\text{H$_2$O}.
\end{equation} 
Other oxygen allotropes (e.g., O$_2$ and O$_3$) have also been investigated in theory and laboratory, showing alternative pathways enriching the solid-state H$_2$O abundance via the same intermediate OH-radicals (see Fig. 4 in \citealt{Linnartz2015}). Meanwhile, it has been suggested that OH-radicals can react with hydrocarbons (e.g., alkynes and methyl radicals) or molecules (e.g., CO) forming alcohols and CO$_2$, respectively \citep{Basiuk2004, Ioppolo2011, Qasim2018}. In general, however, most OH-radicals are expectedly transformed into H$_2$O ice, given abundant (atomic and molecular) hydrogen present in molecular clouds \citep{Cuppen2007}. Astronomical observations toward various cloud-embedded young stellar objects (YSOs) have shown the H$_2$O ice is predominant on dust grains and suggested a thick H$_2$O-rich ice layer (tens of monolayer; \citealt{Boogert2015}). Later, in cold and dense clouds (A$_v>$10, typically within 10$^5$ years assuming n$_\text{H}$=10$^4$ cm$^{-3}$), CO takes over the gas accreting processes forming an apolar ice layer on top of H$_2$O ice and increasing the total thickness of ice mantle by a factor $\sim$2 \citep{Boogert2015}. In the rest time of molecular clouds  (i.e., 10$^{5-7}$ years) before thermal sublimation, several simple molecules, which are preserved in the ice mantle, are exclusively irradiated by energetic particles, such as cosmic rays, UV, and X-ray photons. For example, deeply buried H$_2$O ice covering carbonaceous grain is expected to be dissociated, resulting in OH-radicals and H-atoms. These suprathermal H-atoms and OH-radicals (in the ground or excited states) have been proposed to react with adjacent C$_2$H$_2$ or C$_2$H actively \citep{Michael1979, Smith1984, Senosiain2005, McKee2007}. These induced hydroxylation (or hydrogenation) reactions to C$_2$H$_2$ further augment COM yields in the H$_2$O-rich ice layer in addition to the non-energetic scenario prevailing in earlier translucent clouds \citep{Chuang2020}. Moreover, the studied formation network of C$_2$H$_{\text{n}}$O species starting from C$_2$H$_2$ is also expected to take place at later stages of star formation, such as midplanes of protoplanetary disks or cometary ices, where C$_2$H$_2$ has been abundantly identified in H$_2$O-rich ice \citep{Brooke1996, Altwegg2019}. The chemistry induced by cosmic rays impact is still active at these stages but accounts for different fraction of all energetic inputs.
 
The investigated interactions of C$_2$H$_2$ with H-atoms or OH-radicals, which originate from gas-phase accretion (i.e., non-energetic processing on dust grains) or H$_2$O dissociation (i.e., energetic processing in ice bulks), all lead to qualitatively similar products. In this work, experimental results prove the chemical connections among vinyl alcohol, acetaldehyde, ketene, and ethanol, supporting astronomical observations showing the contemporaneous presence of C$_2$H$_{\text{n}}$O (n=2, 4, and 6) species in various astronomical objects. In particular, ketene, acetaldehyde, and ethanol have been jointly identified toward several star-forming regions, comet C/2014 Q2 (Lovejoy), protostars NGC 7129 FIRS 2, SVS13-A, IRAS 16923-2422A and B, Sgr B2(N2), protostellar L1157-B1 shock regions, and even molecular cloud B5 \citep{Bisschop2007, Fuente2014, Lefloch2017, Taquet2017, Bianchi2018, Biver2019, Jorgensen2020, Manigand2020}. Moreover, acetaldehyde and ketene have been abundantly found in dark and dense clouds, such as B1-b and L1689B, as well as in translucent clouds, such as CB 17, CB 24, and CB 228, implying a low-temperature origin for these species \citep{Turner1999, Bacmann2012, Cernicharo2012}. A quantitative comparison for all these detections shows that the amounts of acetaldehyde, ketene, and ethanol are quite different in early molecular clouds and collapsing stages; ketene and acetaldehyde are more commonly detected in molecular clouds, while ethanol (the hydrogen-saturated species) is generally the dominant molecule in protostellar objects and cometary ices, that is at later evolutionary stages.

Acetaldehyde and the other two isomers (syn-)vinyl alcohol and ethylene oxide, which have a higher energy by 72.4 and 96 kJ mol$^{-1}$, respectively, have all been reported in the massive star-forming region Sgr B2N with a suggested abundance ratio of acetaldehyde: vinyl alcohol: ethylene oxide=800: 1: 1.5, assuming it is an optically thick source \citep{Turner2001}. The uncertainty of the absolute abundance of acetaldehyde has been questioned and remains inconclusive \citep{Ikeda2001}. For low-mass protostars, ethylene oxide has been identified toward IRAS 16923-2422B in the PILS survey\footnote{http://youngstars.nbi.dk/PILS} and towards the prestellar core L1689B, concluding that the abundance of acetaldehyde is at least an order of magnitude higher than the value of ethylene oxide \citep{Lykke2017, Bacmann2019}. An unambiguous detection of vinyl alcohol still lacks in these observations, probably due to relatively weak transitions or chemical instability of vinyl alcohol (enol-keto tautomerization) \citep{Bacmann2019}. Therefore, only upper limits have been reported, namely \textit{N}(CH$_3$CHO)/\textit{N}(CH$_2$CHOH)$\geq$9.2 and 35 for L1689B and IRAS 16923-2422B, respectively \citep{Lykke2017, Bacmann2019}. Also a recent large study strictly searching for vinyl alcohol toward multiple solar-mass protostars has led to setting upper limits \citep{Melosso2019}. These observations consistently point out that acetaldehyde is the most abundant species among the three C$_2$H$_4$O isomers in star-forming regions. These observations are in line with the present laboratory findings that show an efficient chemical transformation of vinyl alcohol to its chemical derivatives such as ketene, acetaldehyde, and ethanol. The backward conversion from acetaldehyde to vinyl alcohol or ethylene oxide is not favorable due to the high internal energy difference.

The relative amounts of ketene, acetaldehyde, and ethanol as a function of energy dose, observed in this laboratory work, is plotted in the mid-panel of Fig. 7 and further compared to the non-energetic experimental findings of similar C$_2$H$_2$ interactions with H-atoms and OH-radicals on dust grains. Upon start of H$^+$ radiolysis, around an energy dose of $\sim$0.5 eV 16u-molecule$^{-1}$, ethanol only contributes to $\sim$4\% of the total yield. Acetaldehyde and ketene are present as main products and account for $\sim$74 and $\sim$22\% of the total yield, respectively. Along with accumulating energy, up to a value of 16.4 eV 16u-molecule$^{-1}$, Fig. 7 clearly shows that hydrogenated species such as ethanol are favored. The composition of ethanol increases by a factor of $\sim$7.3 over $\sim$16.4 eV 16u-molecule$^{-1}$. In contrast, the ketene contribution remains relatively stable (i.e., 22-31\%) before $\sim$4.1 eV 16u-molecule$^{-1}$ due to the competition between formation and destruction mechanisms. Later, it significantly drops when vinyl alcohol is largely reduced. At an energy dose of $\sim$16.4 eV 16u-molecule$^{-1}$, the final composition is found as $\sim$29\% for ethanol, $\sim$63\% for acetaldehyde, and $\sim$8\% for ketene. A very similar chemical trend favoring hydrogen-rich species has been found in previous non-energetic processing of C$_2$H$_2$. As shown in the mid-panel of Fig. 7, the derived composition from the C$_2$H$_2$ + H + OH study is $\sim$51\%, $\sim$41\%, and $\sim$8\% for ethanol, acetaldehyde, and ketene, respectively, after H-atom fluence reaches $\sim$4$\times$10$^{16}$ atoms cm$^{-2}$ under the applied experimental conditions. A higher fraction of ethanol is observed and most likely linked to a larger amount of H-atoms originated from the gas-phase accretion, which enhances the hydrogenation channel of acetaldehyde $\rightarrow$ ethanol. Non-energetic and energetic processing all account for the chemical transformation from unsaturated hydrocarbons to O-bearing COMs under molecular cloud conditions. 

In order to estimate the equivalent time required to accumulate the same amount of energy dose under molecular cloud conditions, an approximation is made by assuming monoenergetic 1 MeV protons with a constant flux of 1 proton cm$^{-2}$s$^{-1}$ prevailing in such regions \citep{Mennella2003, Palumbo2006}. It is important to note that the accumulated energy dose is conventional units, which stand for the effective energy deposited on each molecule with a mass of 16u. Such units is independent of types of impacting ions and the carrying kinetic energies. Therefore, energy deposited on a 16u-molecule originating from 200 keV H$^+$ is considered as the same as the energy dose from a 1 MeV H$^+$ impinging;
\begin{equation}\label{Eq13}
\text{Energy dose$_\text{(200~keV)}$}=\text{Energy dose$_\text{(1~MeV)}$}.
\end{equation} 
According to Eq. \ref{Eq1}, energy dose is the product of stopping power (i.e., \textit{S}) and ion-fluence (i.e., \textit{F}=$\phi\times$\textit{t}, where $\phi$ is the flux in ions cm$^{-2}$s$^{-1}$ and \textit{t} in seconds). Under these assumptions, the 1 MeV proton fluence required to reach the same energy dose in molecular clouds is proportional to the 200 keV H$^+$ fluence in the laboratory. The scaling factors include the ratio of stopping powers (\textit{S$_\text{(200~keV)}$}/\textit{S$_\text{(1~MeV)}$}$\cong$2.78, obtained from SRIM calculation) . Therefore, the estimated time in a molecular cloud can be described by the formula (see \citealt{Palumbo2006} and \citealt{Sicilia2012} for details): 
\begin{equation}\label{Eq14}
\textit{t$_{(1~MeV)}$}=\frac{\textit{S$_{(200~keV)}$}}{\textit{S$_{(1~MeV)}$}}\times\frac{1}{\textit{$\phi_{(1~MeV)}$}}\times\textit{F$_{(200~keV)}$},
\end{equation} 
where $\phi_\text{(1~MeV)}$=1 proton cm$^{-2}$s$^{-1}$. 

The derived corresponding timescale stretches from 1.6$\times$10$^6$ to 5.3$\times$10$^7$ years shown in the upper horizontal-axis by assuming a cosmic ionization rate of $\xi_{CR}$=6 $\times$10$^{-17}$ s$^{-1}$ in molecular clouds. This period is close to the typical lifetime of molecular clouds ($\sim$10$^7$ years; \citealt{Chevance2020}). It is important to note that interstellar ice analogs studied in this work aim to mimic a H$_2$O-rich ice mantle in molecular clouds rather than to represent a realistic ice composition containing several (abundant) ice constituents. Therefore, the obtained experimental data demonstrate the chemical transition of favoring hydrogenated species with energy dose accumulation in star-forming regions. The cosmic ray impact onto the ice mantle additionally enhances the internal conversion of newly formed species taking place during the transition stage from molecular clouds to protostars. The studied solid-state chemistry provides a possible explanation for observational discrepancies between dense molecular cloud (left-handed panel of Fig. 7), such as B5, and later stages (right-handed panel of Fig. 7)), such as low-mass protostellar shock region L1157 B1, protostars IRAS 16293-2422A, B, HH 212, B1-c, and S68N, as well as comet C/2014 Q2(Lovejoy) \citep{Lefloch2017, Taquet2017, Jorgensen2018, Biver2019, Lee2019, vanGelder2020, Manigand2020}, assuming the net transfer from the solid to gas phase is 1:1 ratio for all COMs. The simultaneous and accumulative effects of non-energetic and energetic processing require further astrochemical modeling in order to simulate solid-state reactions occurring on surfaces and in ice bulks containing realistic chemical compositions covering an astronomical timescale. 

Based on the present study on the solid-state chemistry of C$_2$H$_2$ with OH-radicals and H-atoms in the H$^+$ radiolysis of C$_2$H$_2$:H$_2$O ice analogs, the main experimental findings can be summarized as below:

   \begin{enumerate}
        \item C$_2$H$_2$ reacting with H-atoms and OH-radicals, produced by H$^+$ radiolysis of H$_2$O ice, forms (semi-)saturated hydrocarbons, such as C$_2$H$_4$ and C$_2$H$_6$, as well as complex molecules described by the formula C$_2$H$_{\text{n}}$O, such as vinyl alcohol (CH$_2$CHOH), acetaldehyde (CH$_3$CHO), ketene (CH$_2$CO), and ethanol (CH$_3$CH$_2$OH), on grain surfaces at 17 K.

        \item H$^+$ radiolysis experiments on different geometries of C$_2$H$_2$:H$_2$O mixed and layered ices result in qualitatively similar complex molecules, implying that the interface chemistry of H$_2$O ice with unsaturated hydrocarbons on top of dust grains might additionally contribute to COM formation in the solid state. 
        
    \item The experimental results show a kinetic evolution of parent C$_2$H$_2$ and the newly formed complex products as a function of exposure energy; the effective destruction cross-section of C$_2$H$_2$ is (1.9$\pm$0.5)$\times$10$^{-15}$ cm$^2$ per H$^+$ at 200 keV (i.e., 0.07$\pm$0.02 eV 16u-molecule$^{-1}$) and the formation cross-section of the overall two-carbon COMs is (7.2$\pm$0.1)$\times$10$^{-15}$  cm$^2$ per H$^+$ at 200 keV (i.e., 0.26$\pm$0.01 eV/16u-molecule$^{-1}$). The product composition ratio and conversion efficiency are summarized in Table \ref{Table2}.
    
    \item Experimental results suggest that vinyl alcohol, a species carrying a C=C double bond, is the first-generation product in the H$^+$ radiolysis chemistry of C$_2$H$_2$:H$_2$O, acting as a precursor of other O-bearing COMs. The hydrogenation reactions further increase the hydrogen content of the newly formed species by converting ketene to acetaldehyde or acetaldehyde to ethanol, respectively. A clear transition of the chemical composition is found as a function of energy dose, linking the observational compositions toward dense clouds and later star-forming stages. 
   \end{enumerate}

The chemistry of the unsaturated hydrocarbons such as alkyne C$_2$H$_2$ in the current study has been verified in laboratory studies, suggesting an efficient reaction network forming complex organics in interstellar environments \citep{DeMore1969, Hawkins1983,  Hudson1997, Hudson2013, Bennett2005b, Bergner2019, Chuang2020}. In addition to the recombination of single carbon containing radicals, such as HCO, CH$_2$OH, and CH$_3$O on grain surfaces, the investigated ice chemistry between hydrocarbons (containing multiple C-atoms through double or triple bond) and atoms (or radicals) offers an additional formation mechanism leading to molecular complexity in the solid state. A similar mechanism is also expected to form N- or S-bearing COMs. A qualitative study focusing on absolute COM yields for different (non-)energetic processing, such as (atoms) fast ions and electrons, as well as the UV and the X-ray photons, is needed to comprehensively reveal the chemical evolution of these interstellar ices at different stages of star-formation.

\begin{acknowledgements}
This work has been supported by the project PRIN-INAF 2016 The Cradle of Life – GENESIS-SKA (General Conditions in Early Planetary Systems for the rise of
life with SKA). G.F. acknowledges the financial support from the European Union’s Horizon 2020 research and innovation program under the Marie Sklodowska-Curie grant agreement no. 664931 and from Russian Ministry of Science and Higher Education via the State Assignment Contract FEUZ-2020-0038.
Th. H. acknowledges support from the European Research Council under the Horizon 2020 Framework Program via the ERC Advanced Grant Origins 83 24 28. We gratefully acknowledge support by NOVA (the Netherlands Research School for Astronomy),and by NWO, within the framewokr of the Dutch Astrochemistry Network II and a NWO- VICI grant. This work has been supported by the Danish National Research Foundation through the Center of Excellence “InterCat” (Grant agreement no.: DNRF150). 
\end{acknowledgements}

\bibliography{Ref}
\bibliographystyle{aa}




\end{document}